\documentclass{elsart}
\usepackage{graphicx}
\usepackage{dcolumn}
\usepackage{bm}
\usepackage{amsmath}
\usepackage{amsfonts}
\usepackage{amssymb}

\begin{document}
\begin{frontmatter}
\title{Quantum games via search algorithms}
\author{A. Romanelli}
\address{Instituto de F\'{\i}sica, Facultad de Ingenier\'{\i}a\\
Universidad de la Rep\'ublica\\ C.C. 30, C.P. 11000, Montevideo,
Uruguay\\
email: alejo@fing.edu.uy}
\date{\today}
\begin{abstract}
\vspace{0.2cm}

We build new quantum games, similar to the spin flip game, where as a
novelty the players perform measurements on a quantum system associated to a
continuous time search algorithm. The measurements collapse the wave
function into one of the two possible states. These games are characterized
by a continuous space of strategies and the selection of a particular
strategy is determined by the moments when the players measure.
\end{abstract}
\end{frontmatter}

\section{Introduction}

In the field of quantum computation, discoveries like the algorithms of Shor
and Grover \cite{Shor,Grover} showed a superior efficiency with regard to
their classical equivalents. The Shor algorithm factors efficiently any
large number and the Grover algorithm locates a marked item in a disordered
list of $N$ elements in a number of steps proportional to $\sqrt{N}$,
instead of the $O(N)$ of its classical counterpart. This search algorithm
has also a continuous time version \cite{Farhi} that has been described as
the analog analogue of the original Grover algorithm. We have recently
developed a new quantum search algorithm with continuous time \cite{alejo},
that finds a discrete eigenstate of a given Hamiltonian $H_{0}$, if its
eigenenergy is given. The essence of our algorithm, consists on producing a
resonance between the initial state and the looked for state. This resonant
algorithm behaves like Grovers's, and its efficiency depends on the spectral
density of the Hamiltonian $H_{0}$. On the other hand, in the last years,
the theory of games has been used to explore the nature of quantum
information. Initially quantum games were proposed as the quantum
generalization of the classical games but, due to the principles of quantum
mechanics, new game possibilities arose. These studies showed the surprising
result that the quantum strategies could be more efficient than classical
ones. The simplest quantum game, described originally by D. Meyer \cite%
{Meyer}, is the PQ penny flip. It has the virtue of allowing an easy
connection with quantum algorithms. In this game, two players apply
unitary operators alternatingly on the same qubit. The result of the
game depends on the final projection of the qubit over its basic
states. In this context, we propose new quantum games built with our
quantum search algorithm. These games are similar to the PQ penny
flip but we have incorporated to them a continuous time dynamics
together with a quantum measurement process. The strategies of the
players will be determined by the moments when the players make
measurements on the system.

In the following section we review our quantum search algorithm. In section
3 the measurement process is introduced and the dynamical equation of the
system are obtained. In the section 4 we propose different two person zero
sum games and we analyze the phase space of strategies. Conclusions are
drawn in the last section.

\section{Search Algorithm}

\label{sec:resonance} The algorithm is built on a Hamiltonian $H_{0}$ with
normalized eigenstates $\left\{ |n\rangle \right\} $ and eigenvalues $%
\left\{ \varepsilon _{n}\right\} $. Consider a subset \textbf{N} of
$\left\{ |n\rangle \right\} $ formed by $N$ states. Let us call
$|s\rangle $ the unknown searched state in \textbf{N} whose energy
$\varepsilon _{s}$ is given and $|j\rangle $ the known initial
state, whose eigenvalue $\varepsilon _{j}$ is also known. But this
initial state does not belong to the search set \textbf{N}. So
knowing $\varepsilon _{s}$ is equivalent to \textquotedblleft
marking\textquotedblright\ the searched state in Grover's algorithm.
To build the quantum search algorithm a potential $V(t)$ is
necessary that produces the coupling between the initial state and
the searched state. Our proposal \cite{alejo} is the following
potential
\begin{equation}
V(t)=\left|  p\right\rangle \left\langle j\right|  \exp\left(
i\omega _{sj}t\right)  +\left|  j\right\rangle \left\langle p\right|
\exp\left(
-i\omega_{sj}t\right)  \,, \label{potential}%
\end{equation}
where $\left|  p\right\rangle \equiv\frac{1}%
{\sqrt{N}}{\displaystyle\sum\limits_{n\in{\mathbf{N}}}}|n\rangle$ is
an unitary vector which can be interpreted as the average of the set
of vectors in \textbf{N}, and
$\omega_{sj}\equiv\varepsilon_{j}-\varepsilon_{s}$. This proposal
assures that the interaction potential is hermitian, that the
transition probabilities $W_{nj}\equiv\left|  \left\langle
n|V(t)|j\right\rangle \right|  ^{2}=\frac {1}{N}$, from state
$|j\rangle$ to any state of the set \textbf{N} are all equal, and
finally that the sum of the transition probabilities verifies
${\displaystyle\sum\limits_{n\in\mathbf{N}}}W_{nj}=1$. The objective
of the algorithm is to find the eigenvector $|s\rangle $ whose
transition energy from the initial state $|j\rangle $ is the Bohr frequency $%
\omega _{sj}=\varepsilon _{s}-\varepsilon _{j}$, with $\hbar =1$. In order
to perform this task the Schr\"{o}dinger equation, with the Hamiltonian $%
H=H_{0}+V(t)$, is solved. The wave-function, $|\Psi (t)\rangle $, is
expressed as an expansion in the eigenstates $\{|n\rangle \}$ of $H_{0}$, $%
|\Psi (t)\rangle =\sum_{m}a_{m}(t)\exp \left( -i\varepsilon _{m}t\right)
|m\rangle $. The time dependent coefficients $a_{m}(t)$ have initial
conditions $a_{j}(0)=1$, $a_{m}(0)=0$ for all $m\neq j$. After solving the
Schr\"{o}dinger equation the following coefficients are obtained: $%
a_{j}(t)\simeq \cos (\Omega t)$, $a_{s}(t)\simeq \sin (\Omega t)$, $%
a_{n}(t)\simeq 0,$for $n\neq j$ and $n\neq s$ with $\Omega =\frac{1}{\sqrt{N}%
}$. Then the state-probabilities are
\begin{align}
P_{j}(t)& \simeq {\cos ^{2}(\Omega \ t)},  \label{coeficientes} \\
P_{s}(t)& \simeq {\sin ^{2}(\Omega \ t)},  \notag \\
P_{n}(t)& \simeq 0.\text{ }n\neq j\text{ and }n\neq s.  \notag
\end{align}%
From these equations it is clear that a measurement made at the time $t=\tau
\equiv \frac{\pi }{2\Omega }$ has a probability very close to one of
yielding the searched state. This approach is valid as long as all the Bohr
frequencies satisfy $\omega _{nm}\gg \Omega $ and, in this case, our
algorithm behaves qualitatively like Grover's.

\section{Repeated measurements in the algorithm}

Our search has an oscillatory transition between the initial state and the
sought state, the other states are negligibly populated. The wave function
behaves as a time dependent qubit where the coefficients of the eigenstates $%
|j\rangle $ and $|s\rangle $ are alternating in time, that is
\begin{equation}
|\Psi (t)\rangle \approx \ a_{j}(t)\text{ }|j\rangle +a_{s}(t)\text{ }%
|s\rangle \approx \cos (\Omega t)\text{ }|j\rangle +\sin (\Omega t)\text{ }%
|s\rangle ,  \label{qubit}
\end{equation}%
\begin{figure}[h]
\begin{center}
\includegraphics[scale=0.3]{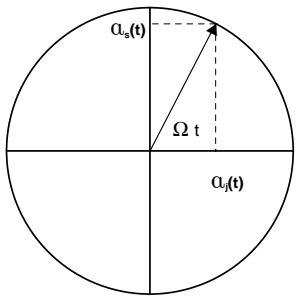}
\end{center}
\caption{{\protect\footnotesize Coefficients of the searched ($a_{s}$) and
the initial ($a_{j}$) states as a function of the time in a circle of unit
radius.}}
\label{game0}
\end{figure}
see Fig.~\ref{game0}. Then the measurement of any observable of the system
produces the collapse of the wave function in the state $|j\rangle $ or $%
|s\rangle $. Supposing that the measurement does not modify the Hamiltonian $%
H_{0}$ and the potential $V(t)$ then, after the measurement, the
resonance algorithm is not affected and continues to work but
starting now from a new initial state, $|j\rangle $ or $|s\rangle $.
Thus, any measurement process will leave the system, with high
probability, in one of these two states. The probabilities
associated with the states $|s\rangle $ and $|j\rangle $ initially
evolve according to the map of eq.(\ref{coeficientes}). If at time
$t_{1}$ the state is measured, the probabilities that the wave
function collapses into either $|s\rangle $ or $|j\rangle $ are
given by the eq.(\ref{coeficientes}) evaluated at $t=t_{1}$.
Immediately after this first measurement, the system is in state
$|s\rangle $ (or $|j\rangle $) and it has an unitary evolution until
the time $t_{2}=t_{1}+\Delta t_{1}$ when a second measurement is
made. \ The probabilities of the states $|s\rangle $ and $|j\rangle
$ after the second measurement at the time $t_{2}$, are given by
\begin{align}
P_{s}(t_{2})& \simeq \cos ^{2}(\Omega \ \Delta t_{1})\ ,  \notag \\
P_{j}(t_{2})& \simeq \sin ^{2}(\Omega \ \Delta t_{1})\ ,  \label{probb}
\end{align}%
(or the same equations exchanging $s$ and $j$). Therefore the system
undergoes an unitary evolution for arbitrary intervals $\Delta
t_{i}=t_{i+1}-t_{i}$ between consecutive measurements and the probabilities
of the states $|s\rangle $ and $|j\rangle $ satisfy the matrix equation
\begin{equation}
\binom{P_{s}(t_{i+1})}{P_{j}(t_{i+1})}=\left[
\begin{array}{cc}
p_{i} & q_{i} \\
q_{i} & p_{i}%
\end{array}%
\right] \binom{P_{s}(t_{i})}{P_{j}(t_{i})},  \label{matrix}
\end{equation}%
where $p_{i}=\cos ^{2}(\Omega \Delta t_{i})$ and $q_{i}=1-p_{i}$ are
transition probabilities, always within the approximation $\omega
_{nm}\gg \Omega $. This last equation looks like a master equation,
which suggests that the global evolution of the system could be a
Markov process. Markovian process have the property that for any set
of successive times ($t_1,t_2,t_3,...,t_n$) the conditional
probability at $t_n$ is uniquely determined by the value of
stochastic variables at $t_{n-1}$ and is not affected by any
knowledge of the values at earlier times. In other words, the system
depends only on the current state and not on the path of the
process. In our case, the measurement of the states of the system is
a simple but extreme form of introducing decoherence that produces a
loss of long range memory. But in eq.~(\ref{matrix}), the matrix of
conditional probabilities is time-interval dependent, then this
equation does not represent a Markovian process. Anyway, a general
solution of the previous equation, for any sequence of measurements,
is obtained.
\begin{equation}
\binom{P_{s}(t_{m})}{P_{j}(t_{m})}=\left[
\begin{array}{cc}
\alpha _{m} & \beta _{m} \\
\beta _{m} & \alpha _{m}%
\end{array}%
\right] \binom{P_{s}(0)}{P_{j}(0)},  \label{solution}
\end{equation}%
where $P_{s}(0)=0,P_{j}(0)=1$ and%
\begin{align}
\alpha _{m}& =\frac{1}{2}\left\{ 1+\prod\limits_{i=0}^{m}\left[ 2p_{i}-1%
\right] \right\}   \notag \\
\beta _{m}& =\frac{1}{2}\left\{ 1-\prod\limits_{i=0}^{m}\left[ 2p_{i}-1%
\right] \right\} .  \label{coefficients}
\end{align}%
If we now consider that the measurement processes are performed at regular
time intervals, $t_{n}=n\Delta t$, eq.~(\ref{coefficients}) becomes
\begin{align}
\alpha _{m}& =\frac{1}{2}\left[ 1+\left( \cos (2\Omega \Delta t)\right) ^{m}%
\right]   \notag \\
\beta _{m}& =\frac{1}{2}\left[ 1-\left( \cos (2\Omega \Delta t)\right) ^{m}%
\right] .  \label{coefficients1}
\end{align}%
In this case all the $\Delta t_{i}$ are equal and the probability
distribution satisfies a master equation then the global evolution,
in a time involving many measurement events, can be described as a
Markovian process. The system has an unitary evolution only between
consecutive measurements. At first sight, for a sufficiently large
number of measurements, the eq.(\ref{coefficients1}) imply that both
$P_{s}$ and $P_{j}$ tend to $1/2 $ independently of the interval
between measurements $\Delta t$ and the initial conditions. However
if the considered total time is fixed the
situation is different. When $m$ measurements are performed in a total time $%
\tau =\frac{\pi }{2\Omega }$, we have $\Delta t=\tau /m$ and the
coefficients are
\begin{align}
\alpha _{m}& =\frac{1}{2}\left[ 1+\left( \cos \frac{\pi }{m}\right) ^{m}%
\right]   \notag \\
\beta _{m}& =\frac{1}{2}\left[ 1-\left( \cos \frac{\pi }{m}\right) ^{m}%
\right] ,\   \label{coefficients2}
\end{align}%
then in this case $P_{s}\simeq 0$ and $P_{j}\simeq 1$ when $m$ $\rightarrow
\infty $.\ This simply means that the more frequently the wave function
collapses, the harder it becomes for the algorithm to significantly depart
from the initial state. Therefore, in this case the algorithm behaves as an
example of the quantum Zeno effect, where the a high frequency of
measurements hinders the departure of the system from its initial state \cite%
{Misra,Chiu}.

\section{The search games}

In above theoretical framework, let us consider a simple quantum state flip
game played between Silvia and Juan. Initially the system is prepared in the
state $|j\rangle $ and the dynamics develops according to the unitary
operator $U(t)$ associated to the Hamiltonian $H(t)$ of the search
algorithm. At the time $T_{1}\in \lbrack 0,\tau )$, Juan measures the state
of the system and afterwards it evolves again with $U(t)$. Silvia knows the
time at which Juan measured but does not know the result of his measurement.
She measures the system at the time $T_{2}\in \lbrack T_{1},\tau ]$ and then
the game concludes. The result of the last measurement determines who wins
the game, if the state is $|s\rangle $ Silvia wins $\$1$ (Juan loses $\$1$)
and if the state is $|j\rangle $ Juan wins $\$1$ (Silvia loses $\$1$).
\begin{figure}[h]
\begin{center}
\includegraphics[scale=0.35]{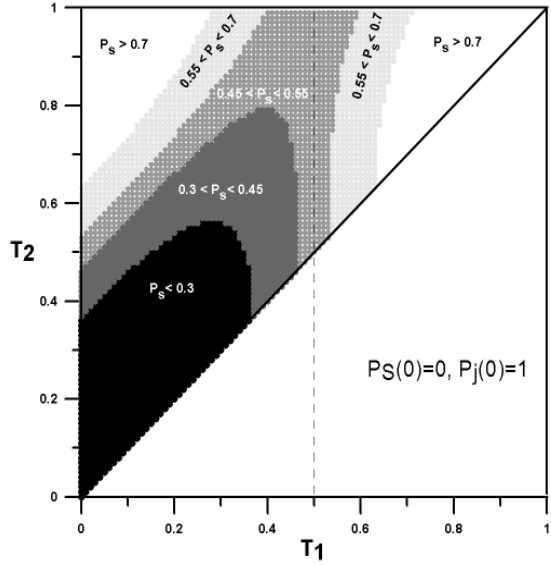}
\end{center}
\caption{{\protect\footnotesize Searched state probability as a function of $%
T_{1}$ and $T_{2}$. The time is expressed in units where $\protect\tau =1.$}
}
\label{game1}
\end{figure}
This is a two-person, zero-sum game, then the payoff to Silvia is the exact
opposite of that to Juan. The players must make their measurements obeying
the condition $0\leq T_{1}\leq T_{2}\leq \tau ,$ but the precise moment when
they measure remains their decision and this determines their strategies.
For example, one of the worst strategies for Juan is to measure at $%
T_{1}\sim \tau ,$ because independently of the time Silvia measures,
he loses almost always. To understand this note that $\tau $ is the
optimal time to obtain the searched state; Silvia's measurement is
very close to Juan's, then the wave function does not evolve (Zeno
effect) and the last state is $|s\rangle $ with high probability.
The probability of the searched state as a function of the times
$T_{1}$ and $T_{2}$ at which Silvia and Juan carry out their
measurements is presented in Fig.~\ref{game1}, where we used the
eq.(\ref{solution}) with $m=2$; its values lie above the straight
line $T_{1}=T_{2}$ due to the condition $T_{1}\leq T_{2}$ and the
different tonalities of gray indicate different probability
intervals. Looking at this figure, the players can plan their
strategies and quickly conclude that it is not an equitable game.
Silvia has many winning strategies, Juan has no winning strategy,
but he has only one strategy that allows him to tie the game. Juan's
optimal strategy is to carry out his measurement at $T_{1}=0.5$
$\tau $, because in this case both players have same probability to
win the game, independently from Silvia's measurement. This could be
an extreme case of the Nash equilibrium \cite{Nash} because any
change in the strategy of Juan could worsen his results and any
change of strategy of Silvia would be indifferent for her. The
existence of this equilibrium adds to the game another element of
interest and in a certain way can surprise us, because we have used
quantum rules in the game (the wave function collapse). But
remembering that the global evolution of this game is a Markovian
process, its existence is a consequence of the mathematical
similarity with classical games \cite{Liu}.

\begin{figure}[h]
\begin{center}
\includegraphics[scale=0.35]{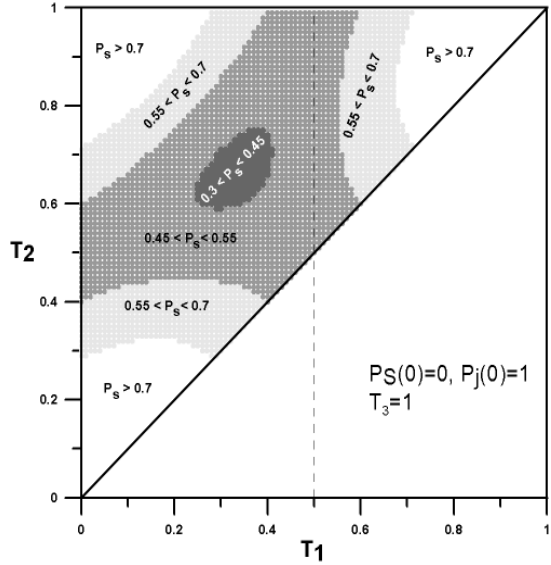}
\end{center}
\caption{{\protect\footnotesize Searched state probability as a function of $%
T_{1}$ and $T_{2}$. The time is expressed in units where $T_{3}=\protect\tau %
=1.$}}
\label{game2}
\end{figure}

An interesting variant of the previous game is obtained when a third
measurement at $t=T_{3}=\tau $ is incorporated. In this new game, after
Silvia's measurement at $t=T_{2}$, the system evolves unitarily until anyone
of the players makes a third measurement at time $t=\tau $. Again Silvia
wins $\$1$ and Juan loses $\$1$, if the result of the last measurement is $%
|s\rangle $, independently of the result of the previous measurements,
otherwise Juan wins $\$1$ and Silvia loses $\$1$. Fig.~\ref{game2}, shows
the probability of the searched state as function of the times $T_{1}$ and $%
T_{2} $ for the new game. In this figure a new area with high probability
appears when the players make their measurements at the beginning of the
game. It is easy to understand this result using the previous ideas of Zeno
effect and optimal time $\tau $ to make the measurement. The Nash
equilibrium is also present for the same strategy as that used in the
previous game.

In this game the zone $P_{s}\leq $ $0,3$ has disappeared as a result of the
increase of asymmetry in the players' strategies, then Silvia's probability
to win is bigger than Juan's. With the aim to quantify the inequity of these
games we now introduce the mean payoff of the game for each player. In the
first place, note that the probabilities of the eq.(\ref{solution}), with $%
m=2$, have a parametric dependencies on $T_{1}$ and $T_{2}$. Then, using eq.(%
\ref{solution}) for the first game, we define the win density for Silvia $%
\sigma _{s}$ and Juan $\sigma _{j}$ as%
\begin{equation}
\binom{\sigma _{s}(T_{1},T_{2})}{\sigma _{j}(T_{1},T_{2})}\equiv \frac{2}{%
\tau ^{2}}\binom{\beta _{2}(T_{1},T_{2})}{\alpha _{2}(T_{1},T_{2})},
\label{densidad}
\end{equation}%
where%
\begin{align}
\alpha _{2}& =\frac{1}{2}\left\{ 1+(2\cos ^{2}(\Omega
(T_{2}-T_{1}))-1)(2\cos ^{2}(\Omega T_{1})-1)\right\} ,  \notag \\
\beta _{2}& =\frac{1}{2}\left\{ 1-(2\cos ^{2}(\Omega (T_{2}-T_{1}))-1)(2\cos
^{2}(\Omega T_{1})-1)\right\} .  \label{alfabeta}
\end{align}%
If the players have as strategy to choose $T_{1}$ and $T_{2}$ at random, but
obeying the constraints, then the probabilities of winning the game are

\begin{eqnarray}
\pi _{s} &=&\int\limits_{0}^{\tau }{dT_{1}}\int\limits_{0}^{T_{1}}{\sigma
_{s}(T_{1},T_{2})}\text{ }{dT_{2}}=0.5  \label{payoffsilvia} \\
\pi _{j} &=&\int\limits_{0}^{\tau }{dT_{1}}\int\limits_{0}^{T_{1}}{\sigma
_{j}(T_{1},T_{2})}\text{ }{dT_{2}=}1-\pi _{s}=0.5,  \label{payoffjuan}
\end{eqnarray}%
for Silvia and Juan respectively. The expected payoff of Silvia is $\overset{%
\_}{\pi }_{s}\equiv \$1\pi _{s}-\$1\pi _{j}$ and the expected payoff of Juan
is $\overset{\_}{\pi }_{j}\equiv \$1\pi _{j}-\$1\pi _{s}$, in this game $%
\overset{\_}{\pi }_{s}=\overset{\_}{\pi }_{j}=\$0$ . In the second game, we
define the win densities as in first game but substituting $\alpha _{2}$ and
$\beta _{2}$ by%
\begin{align}
\alpha _{3}& =\frac{1}{2}\left\{ 1+(2\cos ^{2}(\Omega (\tau
-T_{2}))-1)\right.   \notag \\
& \left. (2\cos ^{2}(\Omega (T_{2}-T_{1}))-1)(2\cos ^{2}(\Omega
T_{1})-1)\right\} ,  \notag \\
\beta _{3}& =\frac{1}{2}\left\{ 1-(2\cos ^{2}(\Omega (\tau
-T_{2}))-1)\right.   \label{alfabeta2} \\
& \left. (2\cos ^{2}(\Omega (T_{2}-T_{1}))-1)(2\cos ^{2}(\Omega
T_{1})-1)\right\} .  \notag
\end{align}%
Now, $\pi _{s}=\frac{7}{8}$, $\pi _{j}=\frac{3}{8}$ and the expected payoff
of Silvia is $\overset{\_}{\pi }_{s}=\$0.5$ and the expected payoff of Juan
is $\overset{\_}{\pi }_{j}=-\$0.5$. Then, they tie in the first game but
Silvia wins a little more in average in the second game, although they have
the same strategy.

Changing the number of measurements allowed in the game and the time
intervals between them it is possible to favor anyone of the players. In the
second game Silvia was the winner but, for example, in games where the
players make many measurements at regular time intervals the eq.(\ref%
{coefficients2}) tell us that Juan will be the winner. Then it is
possible to modify the proposed games or to introduce others using
the equations of the previous section.

\section{Conclusions}

\label{sec:conclusion}

A quantum game, as a quantum algorithm, may be seen as a definite sequence
of unitary transformations acting over a quantum state, in some Hilbert
space. Concepts like interference phenomena, quantum measurements,
resonances, quantum parallelism, amplification techniques, entanglement, etc
should be employed in the new field of quantum games. We have developed a
new kind of quantum game for which there is no classical analogue, it is
simple enough and shows the importance of measurement as\ a fundamental
element in the development of quantum games. These games may be a tool to
study quantum algorithms subjected to external decoherence, as in the
extreme case of measurement \cite{alejo1}.

These games are inspired by the quantum search algorithm but they are very
interesting by themselves. They may be thought, as time dependent games
where to win or to lose is determined by the collapse of the qubit in its
basic states. The games strategies are developed by the players choosing the
times of measurement.

Recent experimental advances allow to obtain and preserve the quantum states
for a system of atoms \cite{Moore}. This opens interesting possibilities to
trap a quantum system with only two energy levels, that allow the
experimental realization of the quantum games proposed in the paper. Finally
we should point out that in this work entanglement is absent, the game is
developed with only one qubit; then a challenge for the future is to
introduce measurements in quantum games with entanglement \cite{abal},
surely in these cases new interesting behaviors will be obtained.

We acknowledge the comments made by Ariel Fern\'{a}ndez and V. Micenmacher
and the support from PEDECIBA and PDT S/C/IF/54/5.


\begin{thebibliography}{99}
\bibitem{Shor} P.W. Shor, in: Proc. of the 35$^{th}$ Annual Symposium on the
Foundations of Computer Science, Ed. S. Goldwasser, Los Alamitos, CA, 1994
and \textit{SIAM J. Comp.}, \textbf{26}, 1484, (1997).

\bibitem{Grover} L.K. Grover, in: Proc. 28$^{th}$ STOC, 212, Philadelphia,
PA (1996); L.K. Grover, \textit{Phys. Rev. Lett.} \textbf{79},325 (1997) M.
Nielssen and I. Chuang, \textit{Quantum Computation and Quantum Information}%
, Cambridge University Press, 2000.

\bibitem{Farhi} E. Farhi and S. Gutmann, \textit{Phys. Rev}. A \textbf{57},
2403 (1998)

\bibitem{alejo} A. Romanelli, A. Auyuanet, R. Donangelo, \textit{Physica}
A, \textbf{360} 274-284 (2006) also in quant-ph/0502161.

\bibitem{Meyer} D.A. Meyer, \textit{Phys. Rev. Lett.} \textbf{82}, 1052
(1999)

\bibitem{Misra} B. Misra and E. C. Sudarshan, \textit{J. Math. Phys.}
\textbf{18,} 756 (1977)

\bibitem{Chiu} C.B. Chiu, E.C. Sudarshan and B. Misra, \textit{Phys. Rev.} D
\textbf{16}, 520 (1977)

\bibitem{Nash} J.F. Nash, \textit{Adv. Math.} \textbf{54}, (1954) 286

\bibitem{Liu} X.F.Liu, C.P.Sun in quant-ph/021204


\bibitem{alejo1} A. Romanelli, A. Auyuanet, R. Donangelo, \textit{Physica}
A, \textbf{375} 133-139 (2007) also in quant-ph/0508142

\bibitem{Moore} F.L. Moore, J.C. Robinson, C.F. Bharucha, B. Sundaram, and
M.G. Raizen, \textit{Phys. Rev. Lett.} \textbf{75}, 4598 (1995)

\bibitem{abal} G. Abal, R. Donangelo, H. Fort, in quant-ph/06071433.
\end{thebibliography}
\end{document}